# *In vitro* Protease Cleavage and Computer Simulations Reveal the HIV-1 Capsid Maturation Pathway

Jiying Ning[1,2,*], Gonca Erdemci-Tandogan[3,*], Ernest L. Yufenyuy[4*], Jef Wagner[3], Benjamin A. Himes[1], Gongpu Zhao[1,2], Christopher Aiken[2,4], Roya Zandi[3] and Peijun Zhang[1,2]

[1]Department of Structural Biology, [2]Pittsburgh Center for HIV Protein Interactions, University of Pittsburgh School of Medicine, Pittsburgh, PA 15260, USA

[3]Department of Physics and Astronomy, University of California, Riverside, California 92521, USA

[4]Department of Pathology, Microbiology and Immunology, Vanderbilt University Medical Center, Nashville, TN 37232, USA

[*]These authors contributed equally to this work

Correspondence: pez7@pitt.edu (P.Z.), roya.zandi@ucr.edu (R.Z.)

**Abstract**

HIV-1 virions assemble as immature particles containing Gag polyproteins that are processed by the viral protease into individual components, resulting in the formation of mature infectious particles. There are two competing models for the process of forming the mature HIV-1 core: the disassembly and *de novo* reassembly model and the non-diffusional displacive model. To study the maturation pathway, we simulate HIV-1 maturation *in vitro* by digesting immature particles and assembled virus-like particles with recombinant HIV-1 protease and monitor the process with biochemical assays and cryoEM structural analysis in parallel. Processing of Gag *in vitro* is accurate and efficient and results in both soluble capsid protein and conical or tubular capsid assemblies, seemingly converted from immature Gag particles. Computer simulations further reveal probable assembly pathways of HIV-1 capsid formation. Combining the experimental data and computer simulations, our results suggest a sequential combination of both displacive and disassembly/reassembly processes for HIV-1 maturation.

**Introduction**

Formation of the infectious Human Immunodeficiency Virus (HIV-1) particle occurs via two processes: the assembly of spherical immature particles that are non-infectious, as the virus buds out at the plasma membrane, followed by maturation of the viral core [1]. During maturation, the viral protease (PR) cleaves the Gag polyprotein into its constituents: matrix (MA), capsid (CA), nucleocapsid (NC), and p6, thereby also releasing the SP1 and SP2 peptides [2]. The interaction between the positively charged NC domain and negatively charged RNA [3], in particular the 5' UTR, is responsible for the encapsidation of the RNA genome within particles. Protein-protein interactions between CA domains are the driving force for Gag assembly in the immature hexagonal lattice [4,5] as well as for CA assembly in the mature capsid [6-8]. Previous computer simulations and theoretical studies have revealed key features of CA self-assembly into conical mature HIV-1 capsids [8-14].

HIV-1 maturation occurs in multiple stages [15]. Following the first cleavage between SP1 and NC, the NC-RNA complex condenses into a dense material. Subsequent cleavage at the MA-CA junction liberates MA and frees CA-SP1 from membrane attachment. The slowest cleavage is the release of SP1 from the C-terminus of CA [15-17]. Proteolytic maturation is essential for infectivity, and PR inhibitors are a key element of current antiretroviral therapies [18]. A potent maturation inhibitor, Bevirimat (BVM), blocks CA-SP1 cleavage and prevents formation of the mature conical capsid [19-22]. Recent structural and mutational studies have indicated that the junction between CA and SP1 could act as a molecular switch to regulate immature Gag assmebly and protease cleavage [23-26]. Structural analyses of the Gag lattice in mutant viruses that have impaired cleavage of Gag at specific sites suggest that processing is ordered and that the RNA/protein complex (RNP) may maintain a link with the remaining Gag lattice after cleavage [27].

While the architectures of immature and mature virions are well characterized [5-8,28,29], the pathway of maturation and the morphological transition process is not well understood. Recent studies have led to two distinct, competing models for the transformation of immature spherical virions to mature virions with conical cores, namely the disassembly/reassembly model and the displacive model [4,27,30-33]. In the disassembly/reassembly model, the immature lattice disassembles following PR cleavage, generating a pool of soluble CA molecules from which a mature capsid assembles *de novo*. Several previous studies provide support for this model. Firstly, based on early cryo-electron tomography (cryoET) observations of 1) virus-like particles (VLPs) with multiple capsids, 2) mature capsids with tip closing defects [34], and 3) a correlation between capsid length and membrane diameter [28], models of capsid growth, from either the "narrow end" [34] or the "wide end" [28] of the capsid, have been proposed. Secondly, structural studies revealed distinct lattice spacings and different interaction interfaces between the immature and the mature CA lattices, suggesting the transition involves a complete disassembly of the immature lattice [4,35]. More recently, cryoET of maturation inhibitor BVM-treated virions displayed a shell that resembles the CA layer of the immature Gag shell but is less complete [21]. Lastly, cryoET of budded viral particles comprising immature, maturation-intermediate, and mature core morphologies suggested that the core assembly pathway involves the formation of a CA sheet that associates with the condensed RNP complex and further polymerizes to produce the mature conical core [33].

By contrast, the displacive transformation model postulates that maturation involves a direct, non-diffusional remodeling of the immature Gag lattice to the mature capsid lattice [30,32]. In this model, the immature lattice does not disassemble to form the mature one. Evidence for a displacive model includes: 1) PR cleavage of *in vitro* assembled CA-SP1-NC tubular assemblies resulted in conversion to mature CA tubes without disassembly [32]; 2) mutant particles with a cleavage defect at the CA-SP1 site have thin-walled spheroidal shells with lattices displacively transformed into a mature-like lattices [31]; and 3) recent cryoET observation of multiple, normal-sized cores within a large membrane enclosure, which argues against *de novo* nucleation and assembly model and suggests a rolling sheet process for CA lattice transformation to conical capsid [30].

To examine the sequence of structural changes that take place during maturation, we establish novel *in vitro* cleavage systems that mimic the maturation process, by digesting purified PR-deficient virions and *in vitro* assembled Gag VLPs with recombinant HIV-1 PR. We further investigate the effects of BVM and Gag cleavage mutants on the *in vitro* maturation process. Using computer simulation, we reveal the impact of membrane and genome on mature capsid formation. Integrating our biochemical and structural findings from the *in vitro* maturation experiments with computer modeling and simulations, we conclude that the HIV-1 maturation pathway is neither simply displacive nor exclusively *de novo* reassembly, but a sequential combination of both displacive and disassembly/reassembly processes.

## Results

### *In vitro* maturation by HIV-1 PR cleavage

To study the structural transitions occurring during HIV-1 maturation, we first established a novel *in vitro* system to mimic the PR-driven HIV-1 maturation process by digesting purified PR-deficient immature virions with recombinant HIV-1 PR. The viral membranes of immature particles were permeabilized with Triton X-100 to allow recombinant HIV-1 PR access to the viral structural polyproteins. The sequential processing of Pr55$^{gag}$ by PR into its constituents in the *in vitro* maturation system closely mimicked that of native virions (Fig. 1). Gel analysis of the PR-treated particles revealed efficient release of MA, CA, and NC in a time-dependent manner, consistent with the process seen in native virion maturation [15]. The PR cleavage was also pH dependent, most efficient at pH 6.0, at which almost all polyproteins were cleaved into CA in 2 hrs at 37°C (Fig. 1a), whereas the majority of CA-SP1 remained intact at pH 7.5. Furthermore, cleavage at the CA-SP1 junction was inhibited by the maturation inhibitor BVM (Fig. 1b), indicating that the immature particles are of the correct structure for binding the inhibitor [36]. Moreover, the MA-CA cleavage site was apparently partially protected by BVM (Fig. 1b), an effect that was not previously observed during HIV-1 maturation [37,38]. The overall cleavage pattern of Pr55$^{gag}$ by PR, observed in this system, suggested that this system could prove useful for mimicking the process of HIV-1 maturation *in vitro*.

In parallel, we established a unique system for *in vitro* maturation of assembled Gag VLPs, using entirely purified recombinant protein components. Immature VLPs were assembled from recombinant Gag protein lacking p6 and most of the MA regions (ΔMA$_{15-100}$Δp6) together with RNA. We tested the cleavage efficiency with various HIV-

1 PR concentrations (Fig. 1c). Under the experimental conditions at pH 6.0, the cleavage process was completed within 2 hrs in the presence of 3.3 µM PR (Fig. 1c), agreeing well with the results obtained from immature virions (Fig. 1a). As with the immature virions, the process was also time-dependent (Fig. 1d); cleavage between CA-SP1 was slow, and all CA was released within 2 hrs in the presence of 3.3 µM PR. Intriguingly, the effect of BVM was pH-dependent, with a strong effect at pH 7.4 but a marginal effect at pH 6.0 (Fig. 1e). The CA-SP1 cleavage process was nearly complete at pH 6.0, despite the presence of a high concentration of BVM (40 µM), whereas a majority of CA-SP1 remained intact at pH 7.4. The same pH-dependent BVM activity pattern was also observed with immature virions. It should be noted that the efficiency of CA-SP1 cleavage was also reduced at pH 7.4 compared to pH 6.0 (Fig. 1e), as seen with immature virions (Fig. 1a). Taken together, these data indicate that both *in vitro* PR cleavage systems recapitulate the HIV-1 maturation process, providing useful tools for characterizing the effects of maturation inhibitors.

**CryoEM analysis of *in vitro* maturation process**

To investigate the structural changes upon HIV-1 PR cleavage, we carried out cryoEM analysis of the *in vitro* maturation process described above. The PR-deficient immature virions appeared as membrane-bound spheres of variable sizes, about 100 nm in diameter (Fig. 2a). The Gag lattices, lining the inside of the membrane, were clearly apparent (Fig. 2a inset, arrow). Limited detergent treatment of immature virions resulted in nonhomogeneous membrane perforation of immature virions (Fig. 2b). Within a single virion, part of the viral membrane became punctured (Fig. 2b and inset, black arrow), while other areas of viral membrane remained intact (Fig. 2b and inset, black arrowhead). In fact, most of the immature particles were broken or fragmented upon mild detergent treatment typical for isolating cores [39] (Fig. 2b). When the concentration of Triton X-100 was reduced to its critical micellar concentration (CMC) of 0.02%, we observed, within membrane enclosed particles, conical structures with an apparent ordered lattice (Fig. 2c), in addition to some ordered densities on the membrane facing outwards that only appear in the protease-treated samples (Fig. 2c, black arrows).

Can such a transition from an immature lattice to a mature lattice take place without membrane and any host cell factors or other viral proteins? To address this question, we employed our all recombinant and reconstituted *in vitro* maturation system and carried out cryoEM analysis of PR-treated Gag VLPs from the same samples analyzed biochemically (Fig. 1c and d, marked with "*"). In the presence of yeast tRNA (10% w/w), purified recombinant Gag polyprotein assembled very efficiently into spherical particles about 100 nm in diameter with well-ordered surface lattices (Fig. 2d and g). The particles were stable and remained intact after 2 hrs incubation with digestion buffer (Fig. 1c "*1", Fig. 2d and g). Upon completion of PR digestion (Fig. 1c "*3"), large clusters, comprising numerous cleavage products, were present (Fig. 2f), in stark contrast to the entirely monodispersed initial Gag VLPs (Fig. 1c "*1", Fig. 2d). Remarkably, the clustered cleavage products formed conical or tubular shaped assemblies, resembling the mature-like structure (Fig. 2i, red arrows), with a cone to cylinder ratio of 57%. Some "hybrid" entities were occasionally observed (Fig. 2i, black arrows, Supplementary Fig. 1), suggesting an incomplete maturation process. We also imaged samples processed with a lower PR concentration (Fig. 1c "*2"). Not surprisingly, we found fewer cones and tubes and more incompletely cleaved particles (Fig. 2e and h, Supplementary Fig. 1), compared to the sample that underwent cleavage to completion. These data suggest that the transition from immature Gag sphere to mature-like conical CA structure can occur without host cell factors, viral membrane or viral proteins other than Gag and PR. Purified CA protein can assemble into cones and tubes *in vitro*, but this requires high protein concentration (>80 μM) and very high salt (>1M NaCl) [8]. Given the relatively low CA concentration (37 μM) and very low salt (near 0 μM NaCl) in our cleavage reactions, *de novo* assembly of cleaved CA into these structures is unlikely. Taken together, these results support, at least in part, a displacive maturation pathway rather than the disassembly and reassembly process [30,32].

**CryoEM structures of Gag VLPs and maturation product**

To confirm that the assembled spherical Gag VLPs do indeed have an immature lattice whereas the final cleavage products have mature-like structures, we obtained 3D density maps of these assemblies by cryoEM. Using cryoET and sub-tomogram averaging [4,5,40], we obtained a 3D density map of Gag VLPs (Fig. 3a) to 18.4 Å resolution by gold-standard Fourier shell correlation (FSC). As shown in Fig. 3b and c, at this resolution, the averaged density map clearly displayed an immature configuration, overlapping well with the previously published immature HIV-1 structure [5]. Docking of the immature structural model 4USN [5] into our density map exhibited a very good fit (Fig. 3b and c), with a cross correlation coefficient of 0.81. Therefore, the assembled Gag VLPs used for *in vitro* maturation are indeed in the immature lattice arrangement.

Upon PR cleavage of Gag VLPs to completion (i.e. all Gag molecules converted to CA, Fig. 1c "*3"), we observed a small fraction of long tubes that were amenable for structural analysis (Fig. 3d), along with numerous conical or short tube-shaped structures (Fig. 2f and i). Combining the segments from the tubes with the same helical symmetry and using a single particle real space helical refinement approach [8,41], we obtained a 3D reconstruction, at 22 Å resolution, of the tubular products that resulted from PR cleavage (Fig 3e-g). In contrast to the undigested precursor, the tube density map showed an entirely mature-like conformation (Fig. 3e), with hexameric CA subunits arranged in helical arrays, as seen in mature CA assemblies [8]. The mature CA structural model 3J34 [8] fitted into the density map very well, further confirming that the tubes have a mature CA conformation. In contrast, docking the mature hexamer 3J34 into the undigested (immature) map and the immature hexamer 4USN into the digested (mature) map resulted in very poor fitting, with cross correlation coefficient values of 0.39 and 0.4, respectively (Supplementary Fig. 2). More interestingly, unlike our previous *in vitro* CA tubular assembly, additional densities, likely corresponding to the released NC and RNA, appeared inside these tubes. A similar NC and nucleic acid release concurrent with formation of mature CA tubes, without CA dissociation, was previously observed upon *in vitro* PR cleavage of CA-NC assemblies [32].

**CA association following Gag cleavage**

We observed only a few conical shaped structures upon *in vitro* cleavage of PR-defective immature virions (Fig. 2c), a surprising result considering the large number of starting immature particles. One plausible explanation for this paucity of cores is dissociation of CA molecules from the particles upon cleavage. To test this, particles were pelleted after PR digestion and particle-associated CA was detected by immunoblotting. As shown in Fig. 4a, a significant amount (more than 1/3) of CA and CA-SP1 remained pelletable, i.e. was associated with the particles or formed aggregates, during the *in vitro* maturation process, after NC and MA were cleaved from Gag. Given that the viral membranes were permeabilized to allow PR to enter viral particles, which would also allow free CA to leave, the particle-associated CA may have arisen from either conversion from immature lattice without dissociation, or via a disassembly/reassembly process. If cleaved CA molecules dissociated from the lattice and reassembled into mature capsid, one might expect from this *de novo* assembly model that capsid-destabilizing mutations and capsid assembly inhibitors would affect the reassembly process and thus inhibit mature capsid formation. We thus investigated the effect of capsid-destabilizing mutation K203A, which does not assemble into tubes in conditions that wild-type does, on CA association following Gag cleavage. As shown in Fig. 4b, the K203A CA mutation did not reduce the fraction of CA that remained pelletable following PR treatment. Similarly, the capsid assembly inhibitor CAP-1 did not inhibit the *in vitro* maturation process (Fig. 4c). These studies suggest that maturation could occur via conversion of the immature lattice to the mature lattice without requiring dissociation of the CA subunits.

On the other hand, *in vitro* cleavage of Gag VLPs yielded very little associated CA in the pellets (Fig. 4d, wt), in contrast to the PR-deficient immature particles (Fig. 4a). In fact, more than 95% of the CA dissociated from the particles and was released into solution (Fig. 4d, wt), consistent with the observed infrequent occurrence of the clustered cleavage products in Fig. 2f and i, seemingly suggestive of a disassembly model. To further investigate the process of *in vitro* PR cleavage of Gag VLPs, we purified Gag proteins carrying substitutions that prevent cleavage at the MA-CA (Y47I mutation in the construct used, or Y132I in the context of full length Gag) and CA-SP1 (L278I/M282I mutations, or L363I/M267I in full length Gag) junctions [42]. Like the corresponding wild type Gag protein, both mutant proteins efficiently assembled into spherical particles. PR treatment of the assembled Gag mutant VLPs yielded the expected cleavage products, MA-CA and CA-SP1, with more than 95% of MA-CA and CA-SP1 in the soluble fraction, as seen for CA in the wild-type VLPs. By contrast, the NC protein released from the wild type Gag protein and both Gag mutants was found in the pellet fraction (Fig. 4d). These studies suggest that the cleavage products, CA, MA-CA and CA-SP1, can dissociate from the particles after removing NC and RNPs.

**Role of membrane and genome in the formation of mature core**

The conversion of spherical immature particles into conical and tubular CA assemblies, together with the observed dissociation of CA molecules from particles upon cleavage, appears to support some aspects of both *de novo* assembly and simple displacive models. We used computer modeling and simulation to explore the underlying physical mechanisms and constraints that are relevant to capsid assembly.

Building upon our previous simulation work [43-46], we used a simple model of viral capsid growth to first investigate the physical impact of the virion membrane on the formation of cylindrical or conical shells. We modeled the membrane as a spherical case, which sets a constraint for the shell growth such that the subunits remain within a sphere of radius, $r_m$. In the model, the membrane-subunit interaction is described as a potential for an interior soft wall, $\sum_i (d_i^m - r_m)^2$, if $r_m < d_i^m$, with $d_i^m$ the distance between the $i^{th}$ vertex of the subunit and the center of the sphere, and the potential vanishes for subunits within the radius $r_m$. As shown in Fig. 5, the membrane restricts the growth of the otherwise extended incomplete shell (Fig. 5a) and induces local stresses causing formation of the pentamers necessary for the assembly of closed conical or tubular shells (Fig. 5b). We note that the limiting effect of membrane on capsid height was reported by Briggs et al. [28]. Their data suggest that the cone assembly is initiated at the narrow tip and stops growing when it reaches the membrane at the wide base. While our results agree in that membrane limits the growth, our simulations indicate a different pathway (see Supplementary Movies 1-6). More interestingly, we found that the membrane could exert a force on the growing sheet that breaks the symmetry and causes the shell to grow as a cone rather than a cylinder (Supplementary Movie 1). Statistical analysis of 300 simulations showed that the frequency of forming a conical shell increased to 52% when constrained in the membrane compared to 37% without the membrane (Supplementary Table 1) suggesting that the virion membrane promotes conical shell formation.

To explore the impact of the genome (or RNP) on shell formation, we modeled the genome as a soft ball, due to steric interactions, with the interaction between the genome and the shell presented through the potential, $\sum_i (d_i^g - r_g)^2$, if $r_g > d_i^g$, with $r_g$ the radius of gyration of the genome (soft ball) and $d_i^g$ the distance between the $i^{th}$ vertex of the subunit and the center of the genome. This potential keeps the subunits outside of the ball. If $r_g < d_i^g$, no interaction occurs between the genome and protein subunits (CA). We emphasize that the focus of our simulations is on the role of RNA-NC complexes in the formation of conical vs. cylindrical shells, and, as such, the soft ball is a good approximation for RNA only after NC cleavage.

We considered the impact of the genome by employing two different models. In the first model, we assumed that a few subunits are attached to the genome but the complexes of the genome-subunits are free to move inside the membrane. If the genome remains attached to the growing sheet during the initial stages of the growth process, the conical shell forms with the genome inside (Fig. 5c, Supplementary Movie 2). If the genome detaches from the growing shell early in the growth process, a cylindrical shell forms with the genome outside (Fig. 5c, Supplementary Movie 3). As shown in Supplementary Table 1 and Supplementary Fig. 7, the presence of the genome increased the frequency of cone appearance to 83%, approximately similar to the previously observed cone frequency of 95% [47].

In the second model, we kept the genome-subunit complexes attached to the membrane during the growth process until about one-sixth of the capsid had formed, at which point, we released the genome-subunit complex. The simulations show that the hexagonal sheet continues to grow asymmetrically along one side of the shell until it meets the membrane, which exerts a force on the shell and makes it roll and grow back towards the RNP (Supplementary Fig. 3, Supplementary Movie 4). Similar to the first model, if the genome is attached to the growing shell, more conical capsids with genome inside occur, while, if the genome detaches, cylindrical capsids form with the genome outside. In both models, the initial attachment of the genome to the growing shell is important for the formation of a conical shell.

**A combination of displacive and disassembly/reassembly pathways**

We further investigated capsid formation from a partially formed shell while in an immature state at the time of release from the membrane. Based on the displacive model, we considered that the sheet of capsid proteins retains the same connectivity as Gag proteins in the immature virion. We assumed that the subunits in the immature sheet, containing as many as 600 subunits, undergo some conformational changes without disassembly resulting in their preferred dihedral angle and flexibility similar to the mature shells. The simulation shows a large sheet of subunits attached to the surface of the spherical membrane and growing. The sheet then detaches from the membrane and subsequently rolls around the genome. Finally, the sheet grows into a closed conical shell in a process identical to the de novo assembly model described above (Fig. 6, Supplementary Movie 5). The outcome of these simulations depended upon the size of the initial immature sheet. For large sheets with around 600 subunits or larger, we observed a decrease in the efficiency of cone formation and an increase in the occurrence of defective cones and cylinders. The sheets with a smaller number of subunits in the initial immature lattice have a much higher probability to grow to cones without defects. When we decreased the number of subunits in the initial "immature" lattice to 400, the lattice rolls around the genome but is not large enough for the two edges to merge immediately. Addition of free CA subunits to the growing sheet results in the formation of cones with almost no defects (Supplementary Fig. 4 and Supplementary Movie 6). In fact, simulation results did not support a direct conversion from the complete immature lattice, as suggested by a simple displacive model. Thus, we conclude that a probable pathway for the formation of conical shells, is a sequential combination of displacive rearrangement and growth. Starting with a partially formed immature sheet has the advantage that the entire lattice does not need to disassemble and reassemble; however, a diffusive growth phase appears to be necessary to minimize assembly defects.

It is important to note that we also carried out a number of simulations in the absence of membrane but in the presence of genome to study the combinatory pathway in situations in which the membrane has been completely removed, as in the case of some experiments. The conical shells were still formed (Supplementary Fig. 5) consistent with the experiments.

**Discussion**

Previous studies have led to two competing HIV-1 maturation models, the *de novo* disassembly-and-reassembly model and the lattice displacive model, for the conversion of immature Gag lattice to mature conical capsid. In an effort to distinguish between these HIV-1 capsid maturation models, we devised *in vitro* PR cleavage systems to mimic HIV-1 maturation and monitored the maturation process with biochemical and structural analysis, as well as computer simulations. The results presented in this study challenge aspects of both existing models. Based on our *in vitro* maturation data and the computer modeling results, we propose a sequential process for maturation in which capsid assembly is initiated by displacive remodeling and capsid growth occurs via soluble subunit addition. The model is illustrated in Figure 7.

Our results suggest that the *de novo* assembly model cannot be the sole pathway, given that *in vitro* PR cleavage of Gag VLPs resulted in conical shaped structures under conditions that do not support assembly of soluble CA molecules into conical or tubular structures [8,47]. Consistent with a recent cryoET study with large membranes enclosing multiple capsids [30], the size of Gag spheres, rather than the viral envelope membrane, which is absent in our *in vitro* Gag VLP system, determines the size of the mature capsid. These data are in conflict with the complete *de novo* disassembly/reassembly model which predicts that the mature core begins to grow at its narrow or wide end, from nucleation of CA units on one face of the membrane, and stops growing when it reaches the boundary of the membrane on the opposite side [28,34]. Furthermore, our study, as well as two recent cryoET analyses [30,33], have identified the formation of CA sheet-like structures as an intermediate during the maturation process. While Woodward *et al.* [33] interpreted this as a *de novo* CA assembly product, data from our and Frank *et al.*'s work [30] suggest that these initial CA sheets form through a displacive process. In addition, HIV-1 particles with functional MA-CA cleavage but defective CA-SP1 cleavage (CA5 particles) exhibit a layer morphology that is consistent with the structure of the mature CA lattice [15,21,27], indicating that the CA-SP1 layer is formed upon release from the membrane and that it is unlikely that most of the CA dissociates from the lattice prior to capsid assembly. Moreover, in our *in vitro* PR cleavage assays, capsid-destabilizing mutations and capsid assembly inhibitors had little effect on CA association, further arguing against the simple *de novo* reassembly model.

On the other hand, our results also argue against an entirely displacive model, as suggested by Frank *et al* [30], given our observations that a major fraction of CA and CA-

SP1 are indeed dissociated from the Gag lattice during *in vitro* PR cleavage of Gag VLPs lacking membrane. If the capsid is formed exclusively by the non-diffusional phase transition of immature to mature CA lattice, one would expect to find that the stress and deformations inherent in this process result in defects in the mature CA lattice. In fact, our computer simulations failed to yield a closed mature core when employing only the non-diffusional phase transition (Supplementary Movie 7), as the transformation of an elastic sheet to a closed conical shell requires disassembly of many subunits at the growing edge and inclusion of pentamers in the "right" position for the shell to close properly. Therefore, based on our experimental data and computer simulation results, we propose a combinatory pathway for HIV-1 capsid maturation, which begins with non-diffusional transition and followed by CA dissociation and growth.

This sequential displacive rearrangement and growth model is consistent with nearly all the previous experimental observations. An assembly pathway involving both displacive and *de novo* modes is advantageous as a part of the shell is already built and does not need to reassemble from scratch. The model further implies that the viral membrane, MA and the RNP likely guide the capsid assembly pathway [33,48]. Interestingly, a recent cryoET study revealed that viral particles with the NC-SP2 domains replaced by a leucine zipper lacked condensed RNPs and contained an increased proportion of aberrant core morphologies [49]. Consistent with our maturation model, a role for HIV-1 integrase (IN) in initiating core morphogenesis and viral RNP incorporation into the mature core has been recently suggested, and IN has emerged as a maturation determinant [50,51]. The effect of IN may explain the apparent paradox that *in vitro* digestion of PR- virions results in 1/3 of CA being pelletable, while there was only ~5% for *in vitro* assembled particles, owing to the presence of Pol in PR- virions and perhaps additional virion-associated host factors. Our model also explains the observed potent trans-dominant effect of uncleavable Gag on viral infectivity, where 4% of Gag with MA-CA linker resulted in a 50% decrease in infectivity [16,52,53]. The presence of even a few such Gag subunits that remain attached to the membrane would preclude CA lattice sheet detachment, thus inhibiting capsid formation and leading to a disproportionate drop in infectivity. Furthermore, in our *in vitro* maturation system, the maturation inhibitor BVM not only blocked cleavage of CA-SP1, but also partially protected the MA-CA cleavage site. The immature lattice-stabilizing effect of BVM has been previously observed in cryoET studies [21] and is reflected in our new combinatory capsid maturation pathway model. Our model also explains the predominance of conical vs. cylindrical structures. We find that any asymmetry developed in the growing lattice due to interaction with the membrane or genome (Fig. 5), or due to the shape of initial immature lattice creates conical capsids, as opposed to cylindrical shells (Supplementary Fig. 6).

While many aspects of the HIV-1 maturation process remain to be explored, our combined biochemical, structural and computational study allows us to integrate our new findings with many previous observations into a new model for HIV-1 maturation. Our model of initial displacive rearrangement from immature Gag followed by growth to the mature capsid, reconciles most of the current experimental evidence and provides a new pathway to elucidate novel therapeutic targets for preventing virion maturation and subsequent infectivity. Moreover, the novel *in vitro* maturation systems that we established in this study will be useful for many researchers to dissect the mechanisms of retrovirus maturation and to test the effects of maturation inhibitors on this process.

**Methods**

**Protein expression and purification**

The cDNA encoding gag polyprotein, *Pr55$^{Gag}$* was obtained from the NIH AIDS Research and Reference Reagent Program, Division of AIDS, NIAID, NIH [54]. To generate Gag (ΔMA$_{15-100}$Δp6), plasmid Prr400 encoding corresponding Gag regions was subcloned into **pET21** (EMD chemicals, Inc. San Diego, CA) using NdeI and XhoI sites. Proteins were expressed in *E. coli*, Rosetta 2 (DE3), cultured in Luria-Bertani media or modified minimal medium, and induced with 0.5mM IPTG at 23°C for 16h. All the mutants were constructed from **pET21** Prr400 with site-direct mutagenesis.

Cell pellets were collected and resuspended in Lysis buffer (25mM Tris, pH7.5, 0.5M NaCl, 1μM ZnSO$_4$, and 10mM 2-Mercaptoethanol) and broken with a microfludizer. Subsequently, 0.11× total volume of 10% polyethyleneimine (pH8.0) was added to precipitate nucleic acids. The lysate was then centrifuged at 35,000 g for 30 min, and the pellet discarded. To the supernatant was added ammonium sulfate to 60% saturation, on ice, with stirring for 2hr. The solution was then centrifuged at 10,000 g for 30 min. The pellet was resuspended in lysis buffer and dialyzed overnight against dialysis buffer (25mM Tris, pH7.5, 50mM NaCl, 1μM ZnSO$_4$, and 10mM 2-Mercaptoethanol). The supernatant was loaded to ion-exchange chromatography (MonoS 10/100 GL column, GE Healthcare, Piscataway, NJ), using 50mM-1M NaCl in 25mM Tris, pH7.5, 1μM ZnSO$_4$, and 10mM 2-Mercaptoethanol. The final purification step for all proteins comprised size-exclusion chromatography (Hi-Load Superdex 200 26/60 column, GE Healthcare, Piscataway, NJ) in 25mM Tris, pH7.5, 0.15M NaCl, 1μM ZnSO$_4$, and 10mM

2-Mercaptoethanol. Proteins were concentrated to 4mg mL$^{-1}$ using Amicon concentrators (Millipore, Billerica, MA), flash-frozen with liquid nitrogen and stored at −80°C.

*In vitro* **Gag VLP assembly**

Protein at 4 mg mL$^{-1}$ in storage buffer was mixed with yeast tRNA at a nucleic acid/protein ratio of 10% (wt/wt), and slowly diluted to 1.5 mg mL$^{-1}$ by dropwise addition of 50 mM sodium acetate (pH6.0), 100 μM ZnSO$_4$, and 5 mM DTT. The mixture was then incubated at 10ºC overnight.

**Isolation of PR-deficient immature virions**

Immature HIV-1 cores were isolated from particles produced by transfection of 293T cells with the PR-inactive HIV-1 proviral clone R9.PR-[55]. Supernatants from transfected 293T cells were filtered through a 0.45 μm pore-size filter to remove cellular debris, and particles were pelleted by ultracentrifugation (120,000 x*g* for 3 hr at 4˚C using Beckman rotor SW32Ti) through a cushion of 20% sucrose in PBS buffer. The viral pellet was resuspended in 250 μL of STE buffer (10 mM Tris-HCl, 100 mM NaCl, 1 mM EDTA, pH 7.4) and incubated on ice for 2 hr prior to ultracentrifugation (120,000 ×*g* for 16 hr at 4°C) through a 0.5 ml layer of 1% Triton X-100 into a 10 ml linear gradient of 30% to 70% sucrose in STE buffer. Immature cores were recovered from fractions 8 and 9 of the gradient, corresponding to the known density of retroviral cores. Aliquots of the immature core-containing fractions were flash frozen in liquid nitrogen and thawed once for use in PR digestion assays. For cryoEM studies, the immature cores were prepared for use without freezing.

### *In vitro* cleavage of immature virions and Gag VLPs

Frozen aliquots of purified cores from PR-deficient particles were thawed, and 50-100 ng of cores was added to 50 µL reactions containing 100mM NaOAc, 100mM NaCl, 1mM EDTA, 1mM DTT, at the indicated pH and concentrations of recombinant HIV-1 PR (kind gift from Dr. Celia Schiffer, University of Massachusetts). Reactions were incubated at 37°C and stopped by the addition of 20 µM of HIV-1 PR inhibitor (Crixivan). A fraction (~66%) of each reaction was removed, diluted to 200 µl in reaction buffer lacking PR, and pelleted by centrifugation (45,000 rpm for 30 min at 4°C in a Beckman TLA-55 rotor). Pellets were dissolved and subjected to 12% SDS-PAGE and immunoblotting with CA-specific mouse monoclonal antibody 183-H12-5C. For cryoEM studies, PR digestion reactions were performed in 200 µl volumes containing 1500 ng of immature cores. For the cryoEM studies shown in Figure 2 panels a-c, PR-deficient HIV-1 particles were concentrated by pelleting through a 20% sucrose cushion without detergent exposure, resuspended in STE buffer, and used in PR digestion reactions in the presence and absence of Triton X-100. Alternatively, PR-deficient immature HIV-1 particles were treated with 0.02% Triton X-100 in PR digestion reactions.

PR digestion experiments with recombinant VLPs were performed by addition of various concentrations of PR to the Gag assembly mixture, and incubated at 37°C for 2hr. For kinetic analysis of Gag cleavage, 3.3 µM of HIV-1 PR was added to the Gag assembly mixture and incubated at 37°C; at different time points, 25 µl of the digestion reaction mixtures was taken out and put on ice to stop the reaction, and then subjected to

SDS-PAGE and cryoEM analysis. The cleavage products were separated by NuPAGE Novex 4-12% Bis-Tris gel (Invitrogen) and visualized by Coomassie blue staining.

**Incubation with maturation inhibitor**

Gag assemblies at different pH values (pH7.4 and pH6.0) were incubated with 40 μM BVM (kind gift of Chin-Ho Chen, Duke University) and 3.3 μM HIV-1 PR at 37 ºC for 2hr. Then 2 μl of the whole sample was mixed with loading buffer, separated by NuPAGE Novex 4-12% Bis-Tris gel (Invitrogen) and visualized by Coomassie blue staining. The same samples were processed for cryoEM analysis.

**CryoEM specimen preparation and data collection**

Gag assemblies with or without PR cleavage were applied (4 μl) to the carbon side of a glow discharged perforated Quantifoil grid (Quantifoil Micro Tools, Jena, Germany). The grids were then manually blotted with a filter paper from the backside, to remove the excess fluid, and plunge-frozen in liquid ethane using a home-made manual gravity plunger. For cryoEM imaging, the frozen grids were loaded into a cryo-holder (Gatan Inc., Pleasanton, CA) and inserted into a Tecnai F20 transmission electron microscope (FEI, Inc., Hillsboro, OR) and imaged with a 2k×2k charge-coupled device camera (Gatan). Low dose (10~20 $e^-/Å^2$) projection images were collected at a nominal magnification of 50,000 with a pixel size of 4.52 Å and underfocus values of approximately 5.0 μm.

**3D reconstruction of helical tubes**

Well-ordered long straight tubes from Gag cleavage samples were Fourier transformed and indexed for helical symmetry. Two tubes belonging to the same helical family (-14, 11) were included in the final reconstruction. In total, 333 segments of 130×130 were boxed out for image processing, and 3D helical refinement and reconstruction were carried out as previously described [41]. During the refinement, helical symmetry and contrast transfer function correction were applied. The UCSF Chimera package was employed for 3D visualization and isosurface rendering [56]. The resolution of the final 3D reconstruction was estimated to be 28/22 Å at the 0.5/0.143 thresholds respectively, from the gold standard Fourier shell correlation (FSC) curve.

**Cryo-electron tomography and sub-tomogram averaging of VLPs**

Gag VLP assemblies (4 µl) were applied to the carbon side of glow discharged perforated R2/2 Quantifoil grids and quickly mixed with 1 µl of a 10 nm fiducial gold bead solution before plunge-freezing using a manual gravity plunger. Tomography tilt series were collected at a nominal magnification of 59,000 × (effective pixel size of 1.9 Å) on a Tecnai G2 Polara electron microscope operated at 300 kV. A series of images were recorded on an FEI Falcon II direct electron detector by tilting the specimen from -45° to 55° in increments of 3° tilt angles. Altogether 34 images were collected in one tilt series with a dose of ~30 $e^- Å^{-2}$. Images were recorded at a defocus value of 4-5 µm using FEI batch tomography software.

Eight tilt series were corrected for the phase inversions due to the Contrast Transfer Function (CTF) using a tile based approach, and subsequently aligned and reconstructed using weighted-back-projection in IMOD [57]. We selected 28 Gag VLPs for further processing from the resulting tomograms. To extract sub-tomograms corresponding to the Gag lattice, initial positions respective to a Cartesian grid defined by each tomogram, were approximated by using a template matching algorithm implemented in Matlab with a reference derived from the previous published immature structure low-pass filtered to 6nm (EMD-2706) [5]. This resolution as well as a coarse angular search were chosen to eliminate any statistical correlation of high resolution information between half data sets in later image processing steps. Following template matching and sub-tomogram extraction, which we limited to the top scoring 300 sub-tomograms from each VLP, the data were randomly split into two groups of ~ 4,200 sub-tomograms each, which were processed independently for all subsequent steps. Each sub-tomogram was iteratively aligned to the average from its respective half-set by using standard cross-correlation based methods. After eliminating the lowest scoring sub-tomograms, the remaining half-sets had ~ 2,000 Gag-hexamers each, with the Fourier Shell Correlation (FSC) between them reaching 23.2/18.4A at the 0.5/0.143 thresholds respectively.

**Computer model building**

Computer simulations were performed based on a model that we had previously developed and employed to successfully explain the *in vitro* formation of conical structures in the absence of membrane and genome as well as the defective structures observed in the *in vivo* experiments [46]. Simulations based upon these previous works were designed to test the impact of the genome and the virion membrane on the formation of conical shells and to investigate the different maturation pathways.

The building blocks of both mature and immature shells were modeled as triangular subunits because they can pack to form hexagonal sheets in flat space [44-46] similar to the Gag and CA molecules. The difference between immature and mature capsids in our model is presented in the mechanical properties of the subunits forming the triangular lattice. It is modeled as an elastic sheet that has an associated bending energy resulting from deforming it away from its preferred curvature, and a stretching energy that comes from distorting the shape of the triangles in the mesh. The bending energy is modeled as simple torsional springs between all neighboring triangles, $E_b = \sum_{<ij>} k_b (1 - \cos(\theta_{ij} - \theta_0))$ with the sum over all neighboring triangles $<ij>$, $k_b$ a torsional spring constant. The quantity $\theta_{ij}$ is the angle between the normal vectors, and $\theta_0$ is the equilibrium angle between the triangles. We can relate $\theta_0$ to the spontaneous radius of curvature $R_0$ for the capsid sheet as $\sin(\theta_0/2) = (12 R_0^2 / b_0^2 - 3)^{-1/2}$. The stretching energy corresponds to a network of springs, $E_s = \sum_i \sum_{a=1}^{3} \frac{k_s}{2} (b_i^a - b_0)^2$ with the sum over all subunits $i$ in the mesh, $b_i^a$ is the length of the $a^{th}$ bond, and $b_0$ the equilibrium bond length.

The equilibrium shape of the capsid sheet is found by allowing the mesh to relax to its minimum energy configuration. This is done numerically using a non-linear conjugate gradient method [58]. The minimum of the total energy of the mesh depends on only two dimensionless constants, the ratio of the linear and torsional spring constants also called the Foppl von Karman number $\bar{\gamma} = \frac{b_0^2 k_s}{k_b}$, and the equilibrium angle between the triangles $\theta_0$. The shape of the final shells formed depends upon the values of $\bar{\gamma}$ and $R_0$, and previous work has explored the parameter space and mapped out the types of shells formed [44,45]. The results we are presenting in this paper correspond to one single $\bar{\gamma}$ and $R_0$ where we obtain cylindrical and conical shells rather than spherical or irregular [46,59]. These two dimensionless parameters define the material properties of the elastic sheet. In the next section we briefly explain our growth model.

**Computer simulations: Nucleation and growth**

The assembly of a shell was modeled through the sequential, irreversible addition of a triangular subunit to an incomplete shell (Supplementary Fig. 8). After each assembly step, the shell is relaxed, assuming that the mean time between additions is much larger than the elastic relaxation time. This corresponds to a physical situation in which each new subunit will be added to the optimal position, *i.e.*, the number of newly formed bonds is maximized and the elastic energy is minimized. Due to the number of bonds formed after addition of each subunit, we consider the process is irreversible [44,45], once a pentamer or hexamer formed, they can no longer dissociate.

During the growth process, if there is a location with 5 triangles attached, two exposed edges can either join to create a pentamer or a new triangle can be inserted to form a hexamer (Supplementary Fig. 8a). As the shell grows and gradually starts to roll, the two distance edges of the shell can become close together. If the distance between the edges is very small, they merge to mimic the hydrophobic interaction between the subunits at the edge (Supplementary Movie 1).

**Data availability:**

**Accession codes:** CryoEM structural data have been deposited at the EMDB under accession code EMD-8403 and EMD-8404 for Gag VLPs before and after protease cleavage, respectively.

**Author contributions:**

This study was conceived by P.Z., R.Z. and C.A.  C.A. and E.L.Y. designed and performed *in vitro* cleavage of PR-deficient virions. P.Z. designed the approaches for *in vitro* cleavage of Gag VLPs. J.N. made Gag mutant constructs, purified wild type and mutant Gag proteins, and carried out Gag assembly and *in vitro* cleavage experiments. B.H. and G.Z. performed structural analysis of immature Gag VLPs.  R.Z., J.W. and G.E.-T. designed the simulations. G.E.-T. and J.W. performed the simulations. R.Z. and G.E.-T. analyzed the data. P.Z., R.Z. and C.A. developed the maturation pathway model. P.Z., R.Z. and C.A. wrote the paper with support from all the authors.

**Acknowledgements:**

We thank Dr. Celia Schiffer, University of Massachusetts for the purified recombinant HIV-1 PR protein, Dr. Jing Zhou, Vanderbilt University for virus preps, and Chin-Ho Chen, Duke University, for BVM. We thank Dr. Xiaofeng Fu, University of Pittsburgh, for help with cryoEM imaging and Dr. Teresa Brosenitsch for critical reading of the manuscript. Monoclonal antibody 193-H12-5C was obtained from Dr. Bruce Chesebro via the NIH AIDS Research and Reference Reagent Program. This work was supported by the National Institutes of Health NIGMS Grant P50GM082251, the Office of the Director Grant S10OD019995, the Childhood Infections Research Program (NIH T32 AI095202), and by the National Science Foundation through Grant No. DMR-1310687.

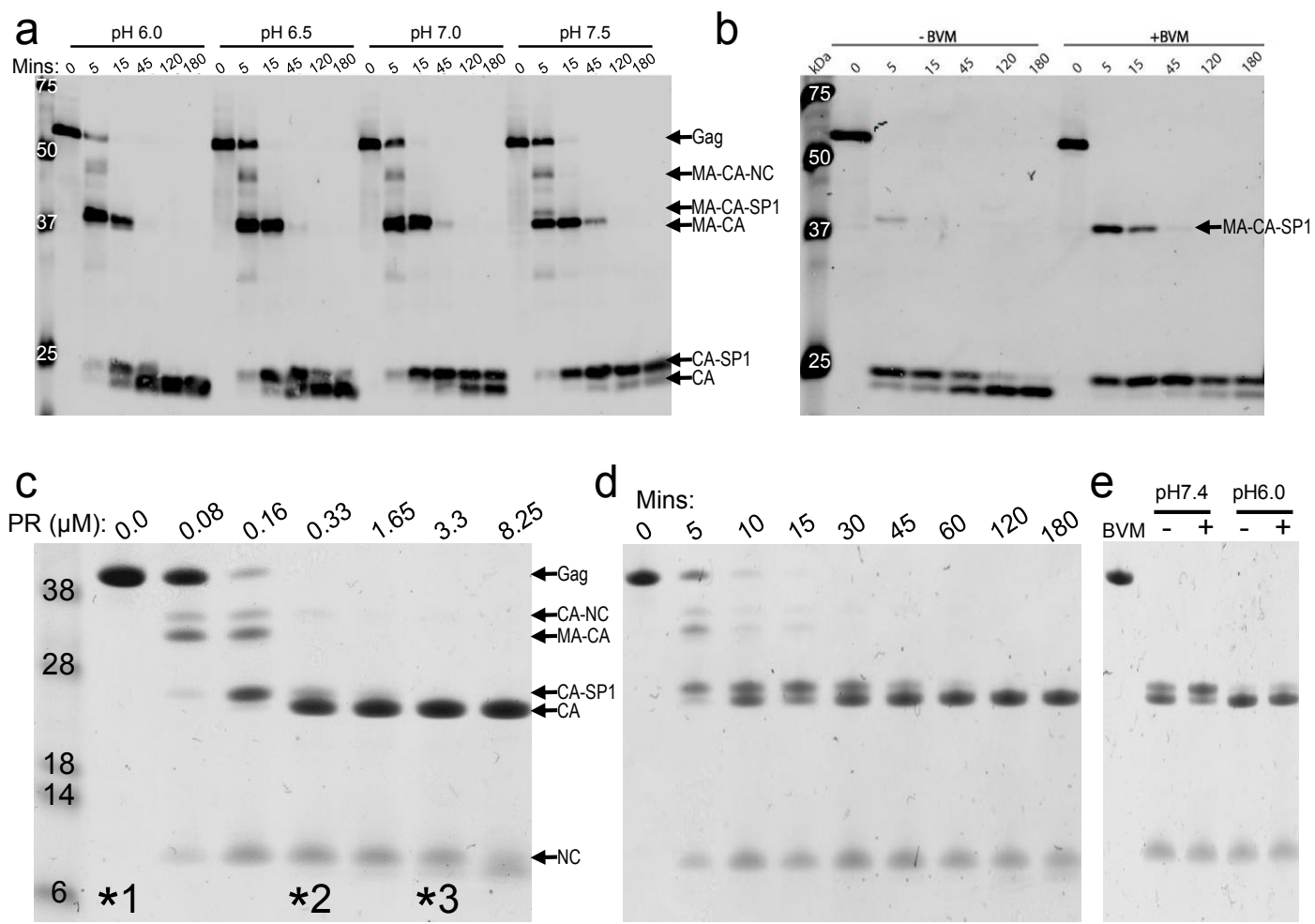

**Figure 1.** *In vitro* maturation of HIV-1 by protease (PR) cleavage. (A&B) Digestion of immature HIV-1 particles with recombinant HIV-1 PR at 37° C. (A) Reactions were performed with 1 µM PR at the indicated pHs. (B) Reactions were performed at pH 7.4 with 4 µM PR the presence or absence of 2 µM bevirimat (BVM). Protein products were immunoblotted with a CA-specific antibody and the bands labeled accordingly. (C-E) Cleavage of *in vitro* assembled Gag VLPs at pH 6.0, at various PR concentrations for 2 hr (C), different times of incubation with 3.3 µM PR (D), and in the presence or absence of BVM (40 µM) at pH 7.4 and pH 6.0 with 3.3 µM PR for 2 hr (E). Protein products were visualized by Coomassie blue staining and are labeled accordingly. Molecular weight markers are shown on the left. The protein designated "MA" contains a large internal deletion (15-100) in Gag VLPs. Samples labeled with "*" were subjected to cryoEM analysis, as shown in Fig. 2.

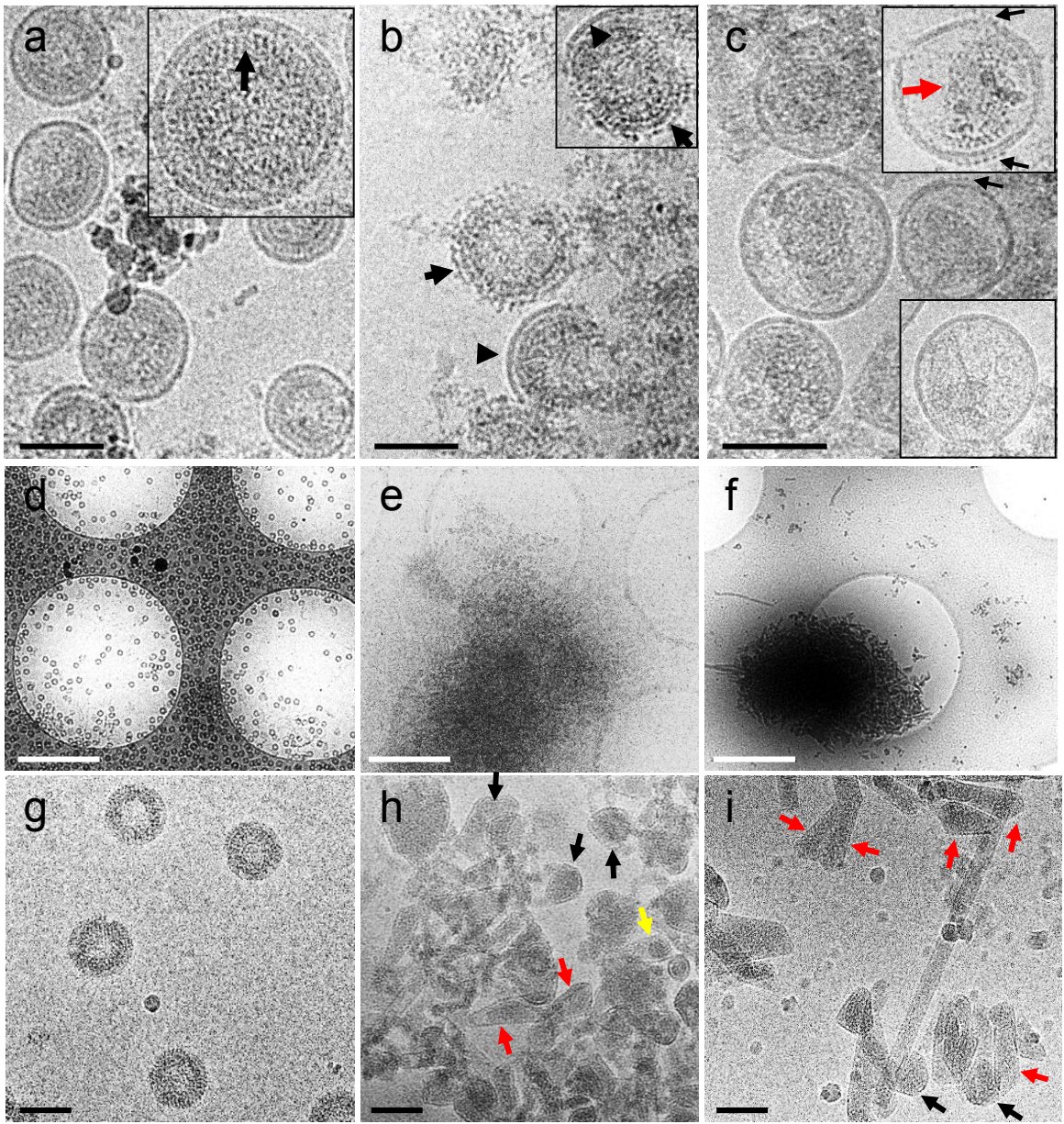

**Figure 2.** CryoEM analysis of HIV *in vitro* maturation by PR cleavage. (A-C) CryoEM images of PR-deficient HIV-1 immature virions before Triton X-100 treatment (A), after Triton X-100 treatment without HIV-1 PR (B), and after Triton X-100 treatment but with PR at 3.3 µM, pH 6 for 2hrs (C). The inset in (A) shows an enlarged view of immature virion, with an arrow pointing to the immature Gag lattice portion. The inset in (B) shows a Triton X-100 treated virion in which part of the membrane was still intact and part was punctured. Intact and perforated virial membranes are labeled with black arrowheads and black arrows in B, respectively. The inset in (C) shows an enlarged view of a virion after PR cleavage, with a red arrow pointing to the ordered lattice of the cleavage product. (D-I) CryoEM images of *in vitro* maturation of Gag VLPs. Low (D-F) and high (G-I) magnification cryoEM images show Gag VLPs morphology changes from spheres (D&G) to cones and tubes after 0.33µM (E&H) or 3.3µM (F&I) PR treatment for 2 hours. Scale bars, 100 nm in A, C and G, 1 µm in D.

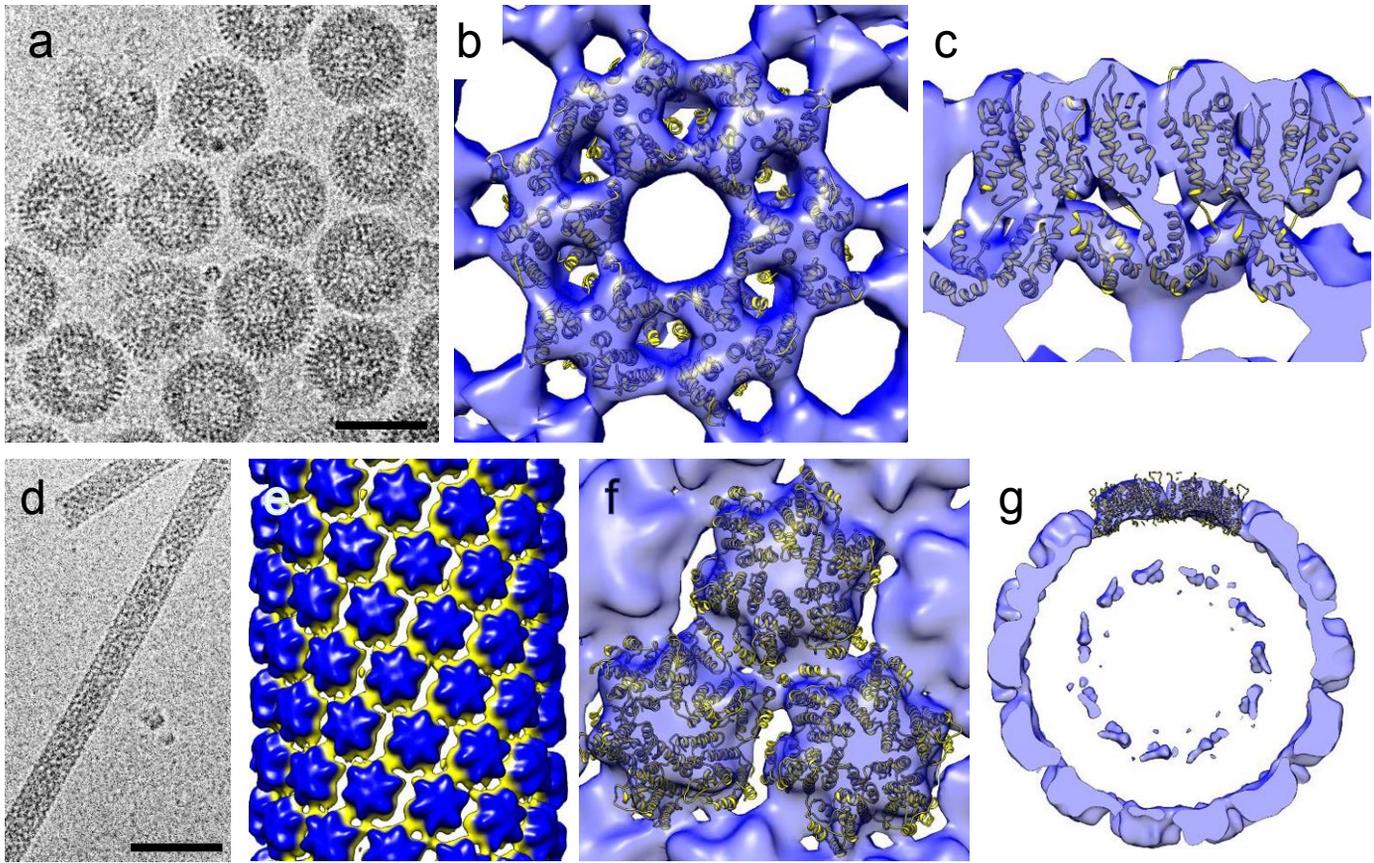

**Figure 3.** 3D reconstruction of Gag VLPs before and after PR cleavage. (A-C) The cryoEM image (A) and the subtomogram averaged 3D map (B and C) of Gag VLPs before PR cleavage. The resulting map is contoured at 1.5σ, with immature CA lattice (PDB code 4USN, yellow) docked into the density map, viewed along (B) and perpendicular to (C) the radial direction. (D&E) The cryoEM image and 3D reconstruction of the tubular structures present after PR cleavage of VLPs. The resulting map is contoured at 1.5σ, and colored radially from yellow to blue. (F&G) Fitting of mature CA hexamer (PDB code 3J34, yellow) into the density map, viewed from tube surface (F) and along the tube axis (G). The map is contoured at 1.5σ. The mature CA hexamer, not the immature, docks well into the EM density map. The inner layer density (seen in G) is likely the cleaved NC protein. Scale bars, 100 nm.

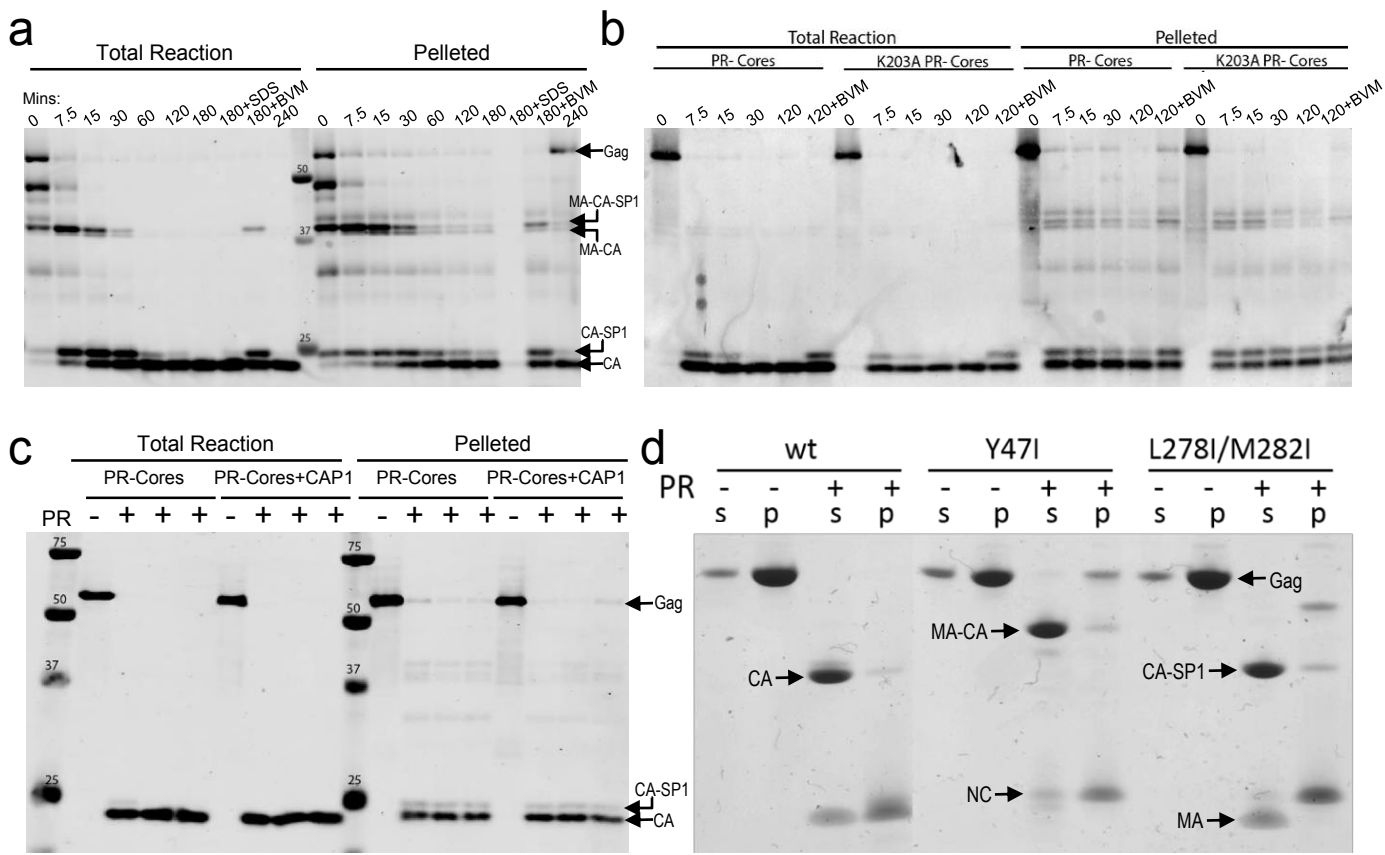

**Figure 4.** CA association following Gag cleavage by PR. (A) Immunoblot analysis of PR-treated immature virions in the presence of bevirimat (BVM) where indicated. PR-negative immature virions were digested with 1 µM PR at 37°C for varying amounts of time, up to 4 hrs. Particles were pelleted after digestion and CA association with core was monitored by immunoblotting with CA-specific mouse monoclonal antibody. (B) Effect of the capsid-destabilizing mutation K203A on CA association following Gag cleavage. Purified cores from PR-defective particles with wild type (wt) or K203A mutant CA were treated and analyzed as in (A). (C) Effect of CAP-1 on CA association during in vitro maturation. Cores from PR-deficient HIV-1 particles were treated as in (A) in the presence of 20 µM capsid assembly inhibitor CAP-1 for 2.5 hrs, then analyzed as in (A). (D) Effect of cleavage mutants on CA association during *in vitro* maturation by PR digestion of Gag VLPs. Supernatant (s) and pellet (p) of cleavage products following 3.3 µM PR digestion at 37°C for 2hrs, analyzed by SDS-PAGE gel and visualized by Coomassie Blue staining. The corresponding cleavage products are labeled.

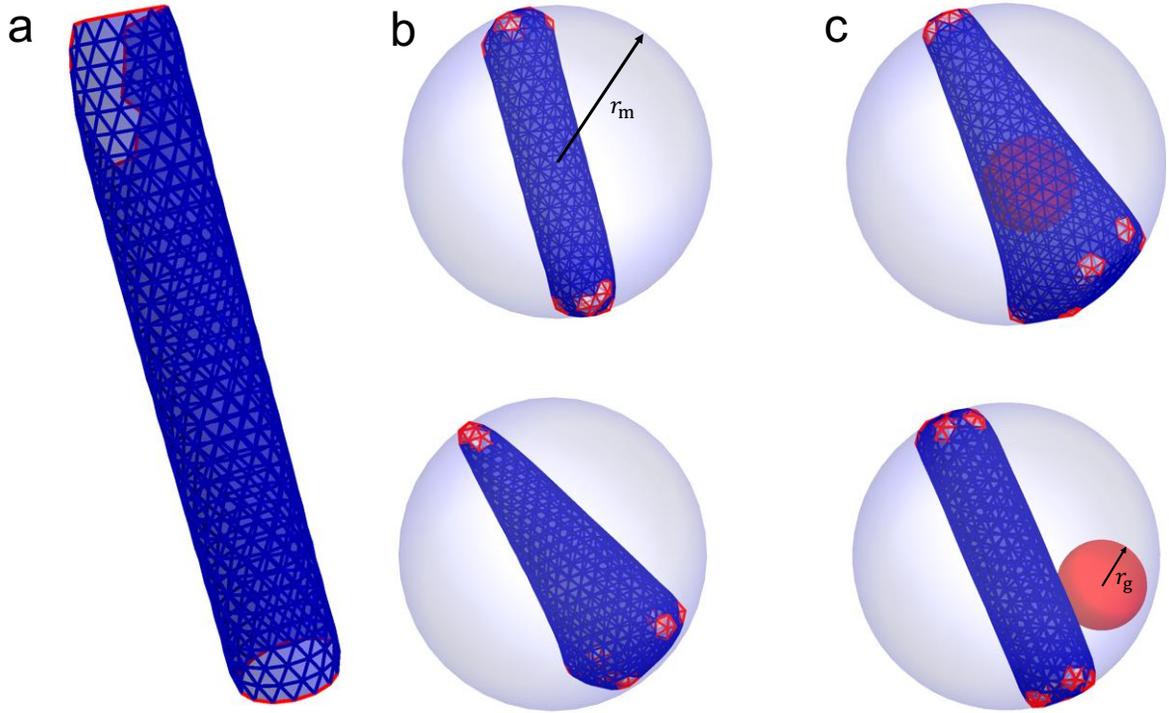

**Figure 5.** Role of membrane and genome in the formation of mature core. (a) Formation of open tube in a free environment. (b) The presence of membrane (the grey sphere) limits the size of the cylinder. Membrane can also facilitate formation of pentamers resulting into the formation of cones. (c) During the assembly process, if the genome (ball) remains attached to a few subunits, a conical capsid forms. Otherwise, a cylindrical shell assembles with genome remaining outside, see the movie in Supplementary Material.

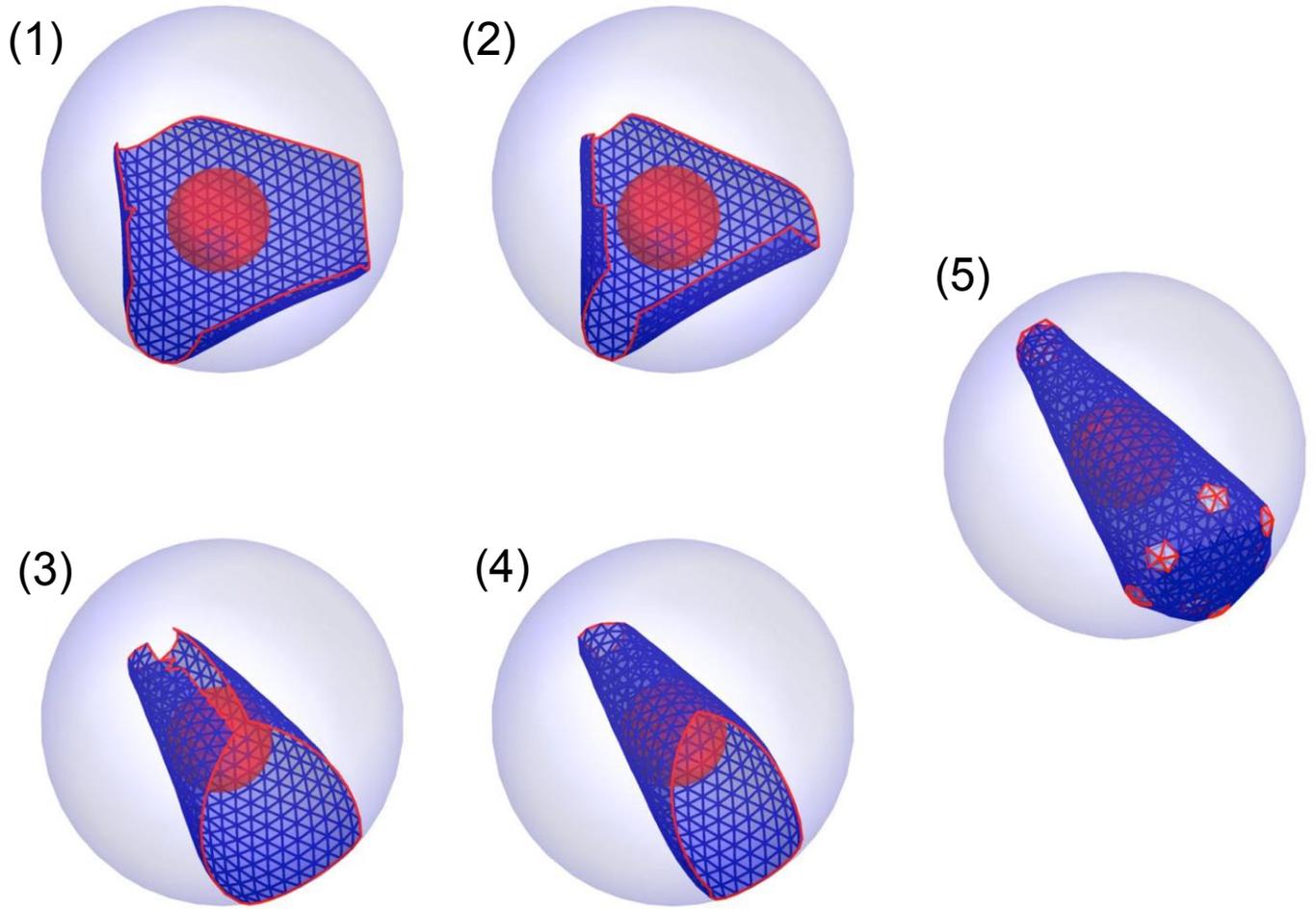

**Figure 6.** Formation of a conical capsid through a combination of displacive and disassembly/reassembly models. Initially a lattice formed from 600 subunits detaches from the membrane. The lattice undergoes maturation (steps 1-5), i.e., the mechanical properties and the spontaneous curvature of lattice transforms from an immature capsid to a mature one.

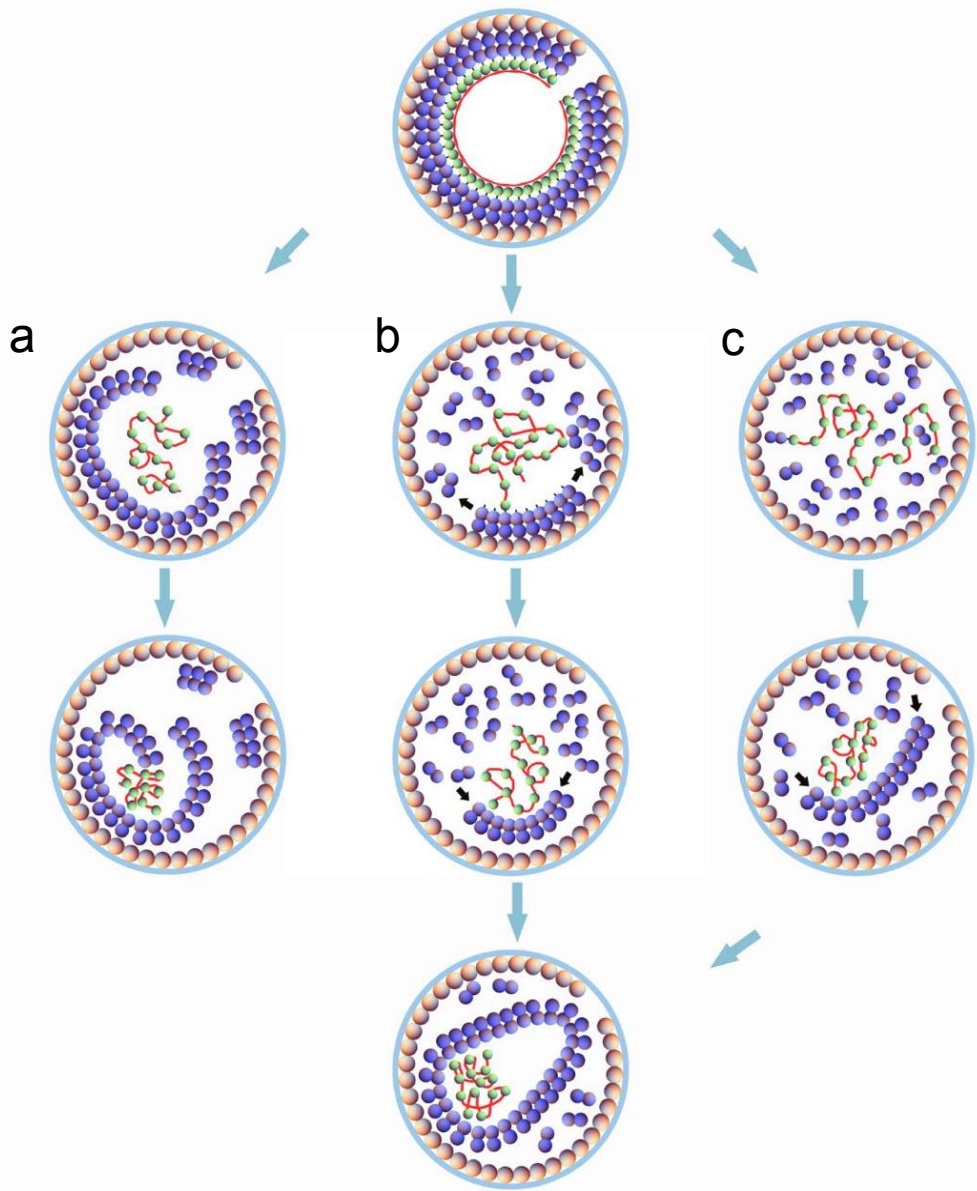

**Figure 7**. A model for different HIV-1 maturation pathways. Top figure shows an immature shell composed of a membrane (blue circle), MA proteins (orange balls), CA proteins (purple balls), NC proteins (green balls) and RNA (the red line). Column A shows the fully displacive model in which the detached immature lattice undergoes maturation. Column B illustrates the sequential combination of displacive and assembly model. A portion of the immature lattice undergoes maturation followed by addition of CA proteins to the growing edge. Column C indicates the disassembly and *de novo* reassembly model. The immature lattice completely disassembles and then the disassembled CA proteins assemble *de novo*.

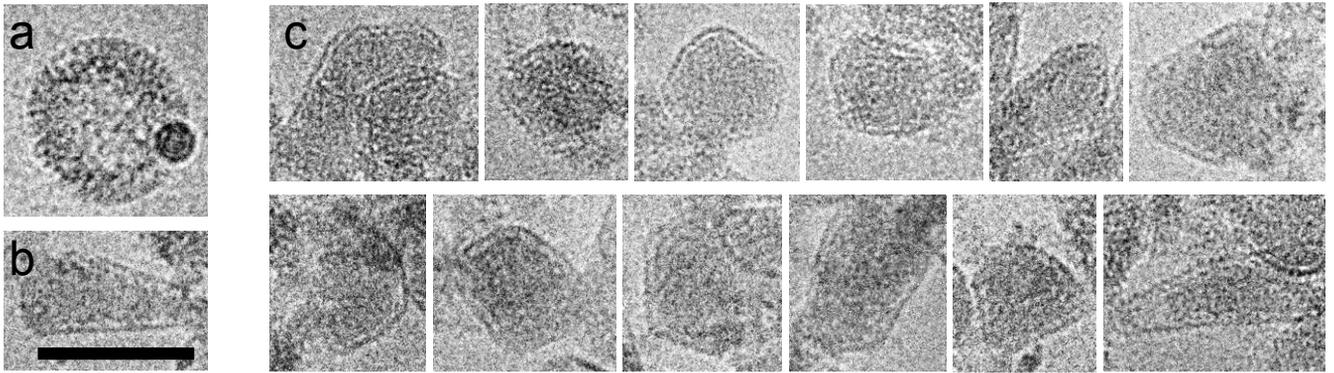

**Supplementary Figure 1.** Maturation intermediates by *in vitro* protease cleavage of Gag VLPs. (a) A typical Gag VLP before PR cleavage. (b) Formation of the mature core after PR treatment. (c) A gallery of "hybrid" entities resulting from incomplete maturation process. Shown are the changes from spherical immature surfaces to sharply curved surfaces. The images were recorded from the same sample in Fig. 1c labeled "*2". All panels are on the same scale. Scale bar, 100 nm.

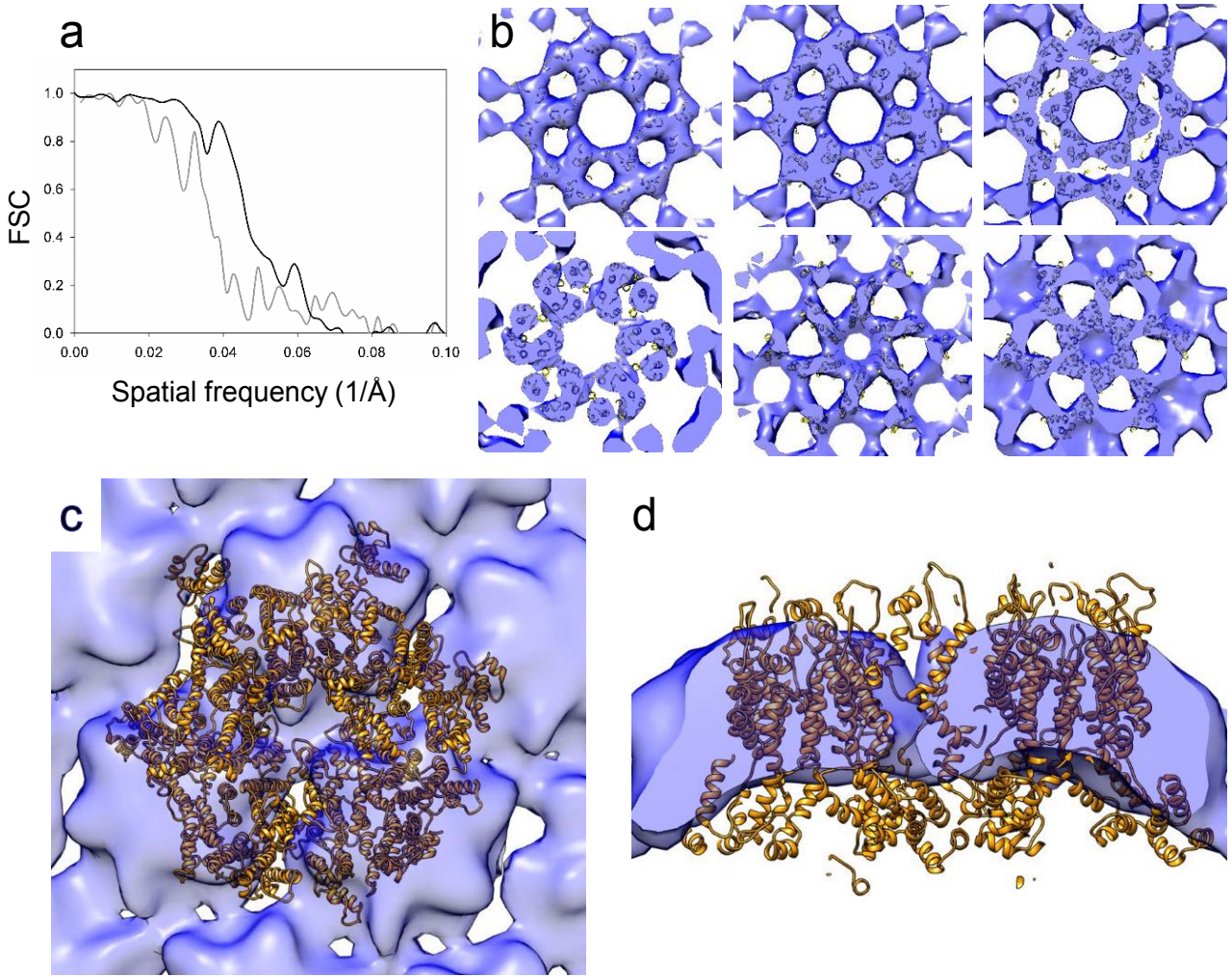

**Supplementary Figure 2.** (a) Fourier Shell Correlation (FSC) of the 3D density maps of Gag spheres (black) and post-cleavage tubular products (grey). (b) Fitting of immature CA lattice (PDB code 4USN) into the Gag sphere density map, shown in slice-views along the radial direction. (C&D) Fitting of immature CA lattice (PDB code 4USN) into the tube density map, viewed from tube surface (c) and along the tube axis (d). The maps are contoured at 1.5 σ.

(1) 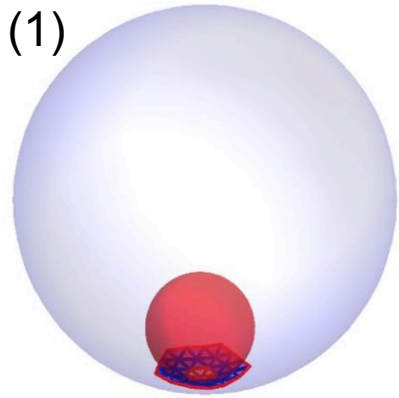
(2) 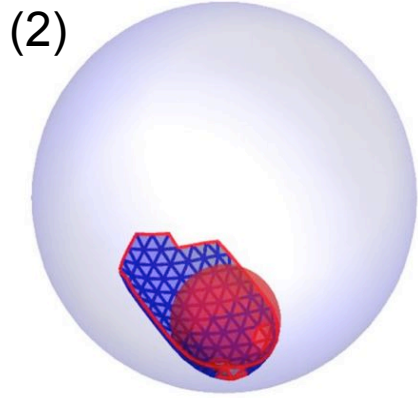
(3) 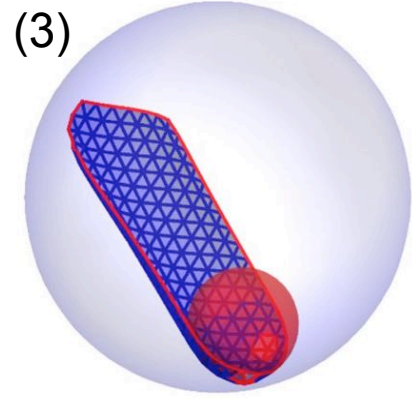
(4) 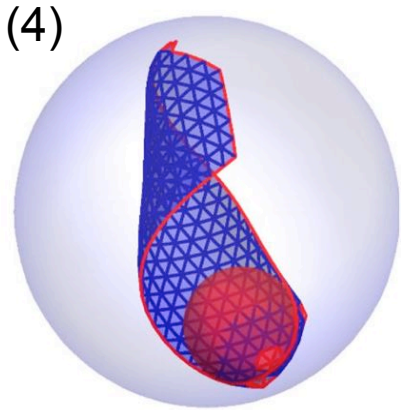
(5) 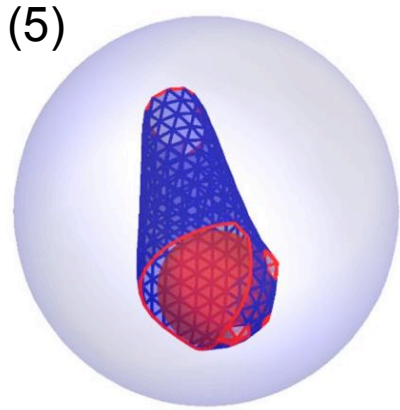
(6) 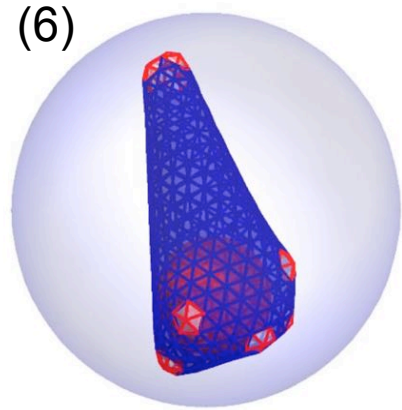

**Supplementary Figure 3.** Formation of a conical capsid. The genome and a few subunits remain attached to the membrane at the beginning of the growth process.

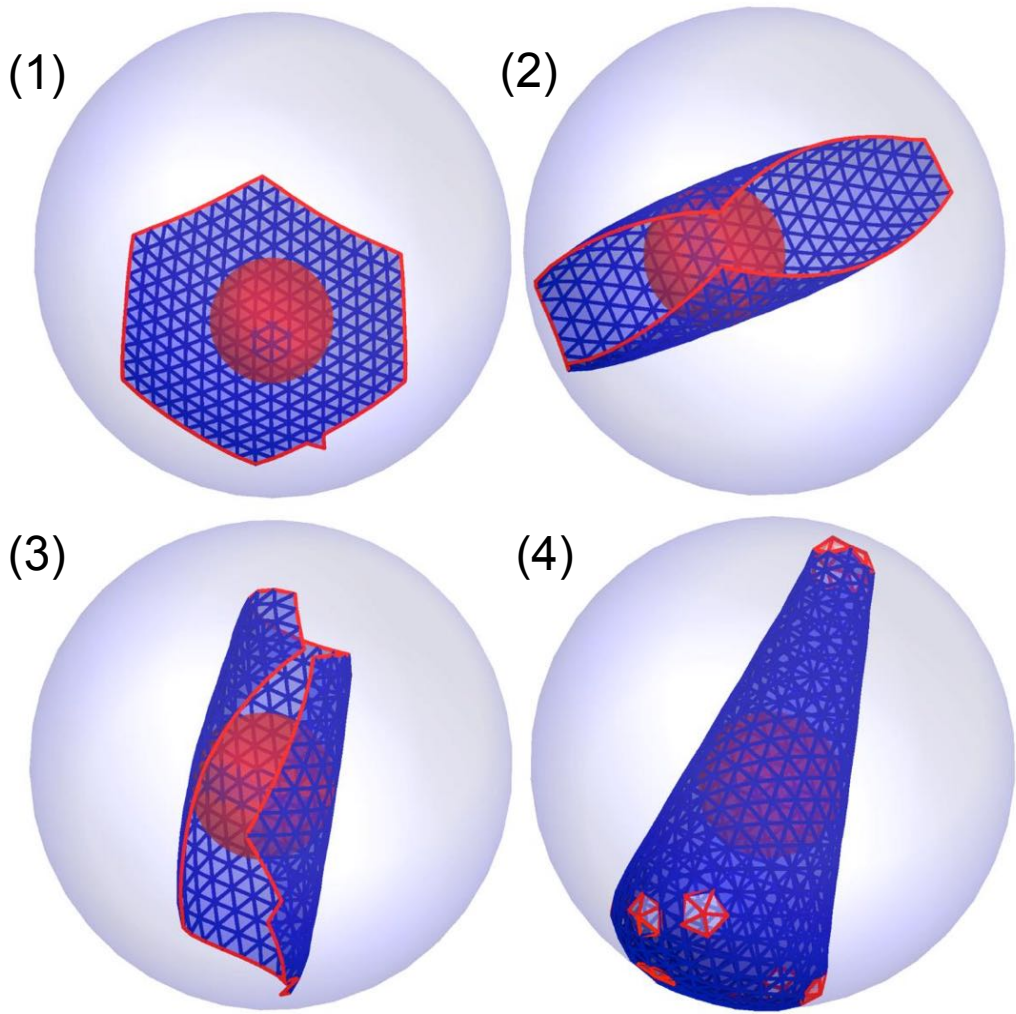

**Supplementary Figure 4.** Snapshots of formation of a conical capsid through a combination of displacive and disassembly/reassembly models. Initially a lattice formed from 400 hundred subunits detaches from the membrane. The lattice undergoes maturation, i.e., the mechanical properties and the spontaneous curvature of lattice transforms from an immature capsid to a mature one. The orange ball illustrates the genome.

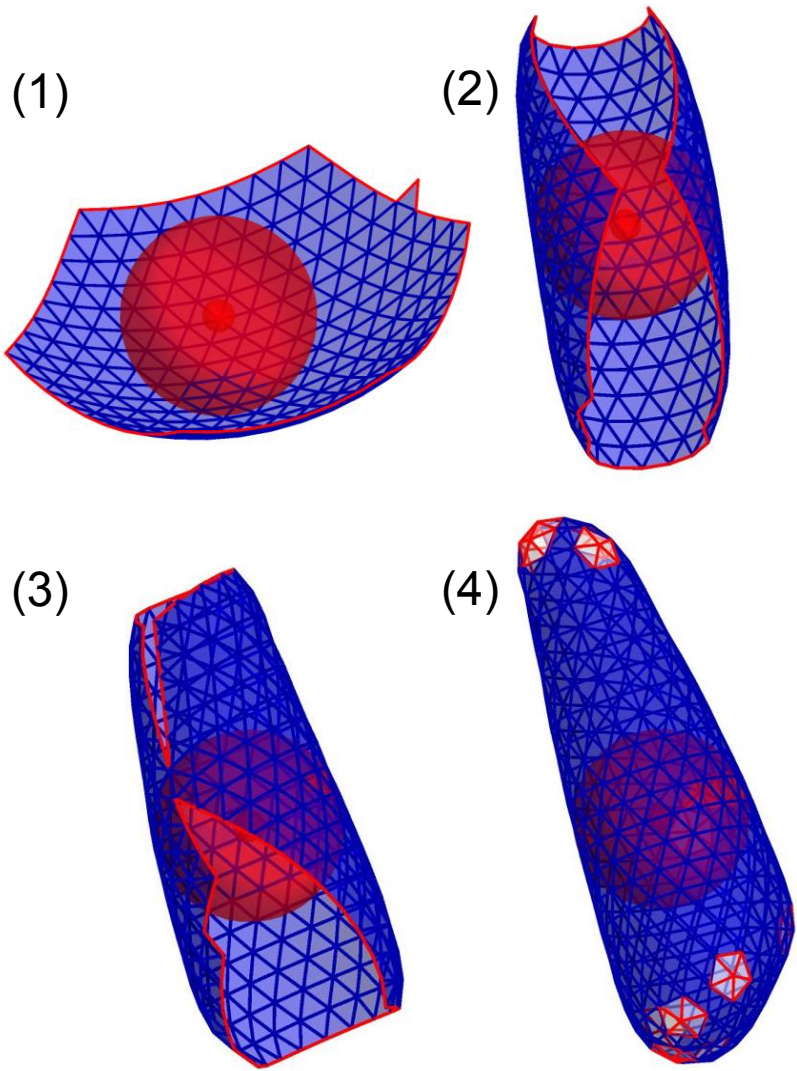

**Supplementary Figure 5.** Snapshots of formation of a conical capsid through a combination of displacive and disassembly/reassembly models in the absence of membrane but the presence of genome.

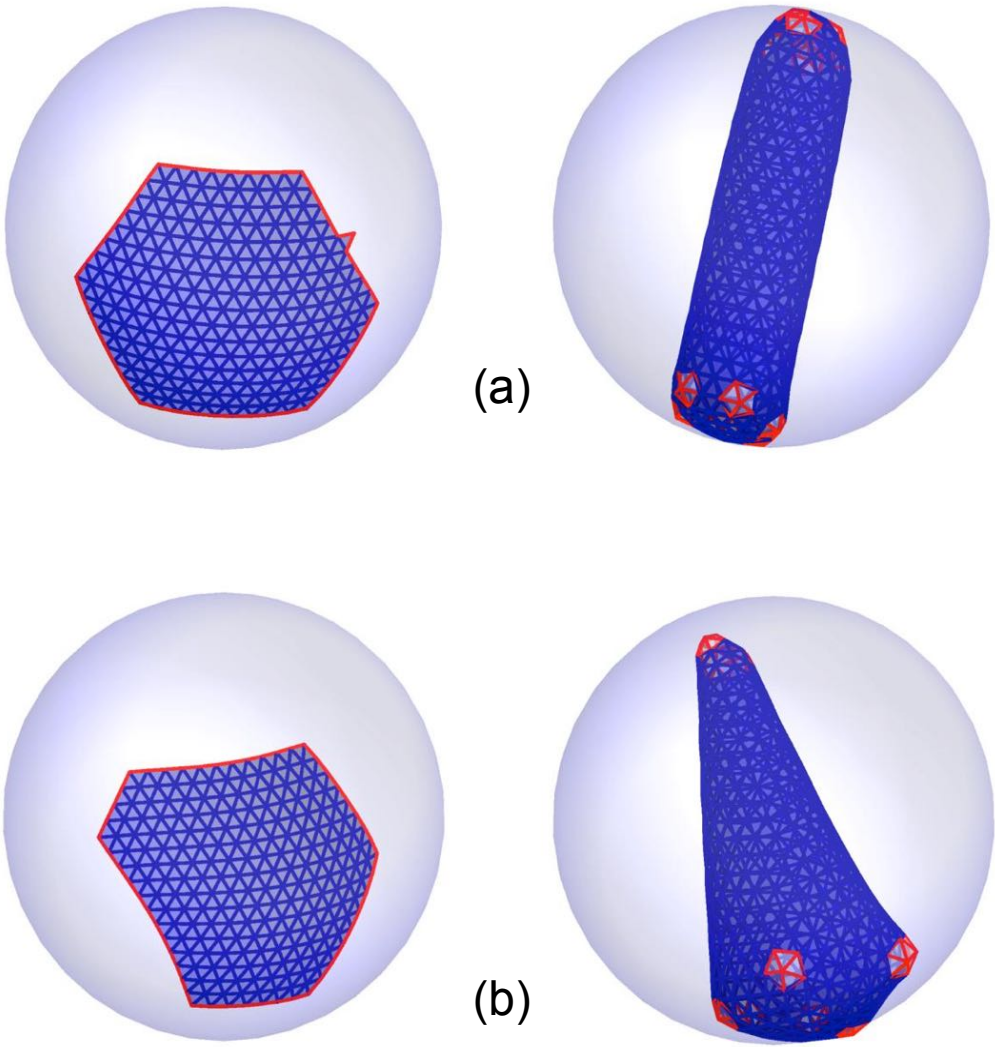

**Supplementary Figure 6**. The role of initial condition and symmetry in the final structure of capsids. If the initial lattice has hexagonal symmetry, a cylindrical shell forms (a); otherwise a conical capsid forms (b).

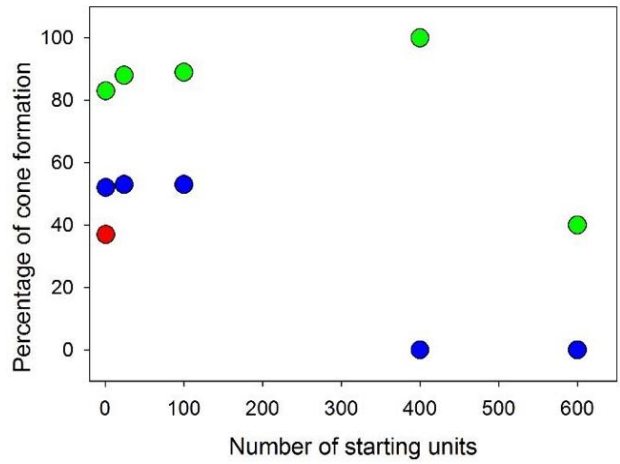

**Supplementary Figure 7**. Relative frequency of cone formation in simulations under the indicated conditions: in the absence of membrane and genome (red); in the presence of membrane but absence of genome (blue); and in the presence of both membrane and genome (green).

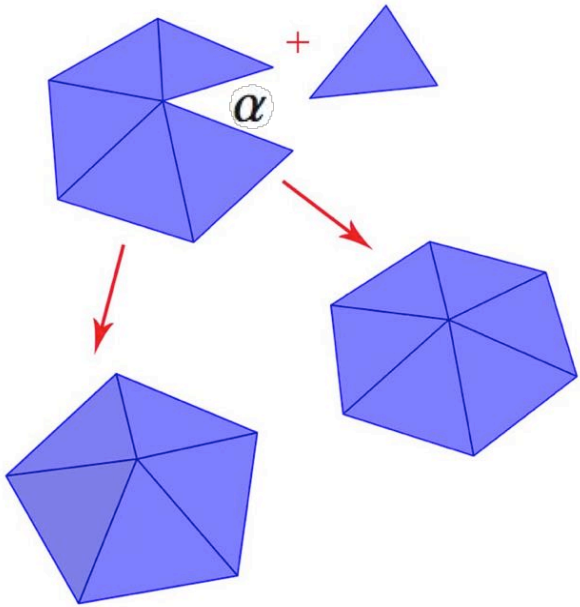
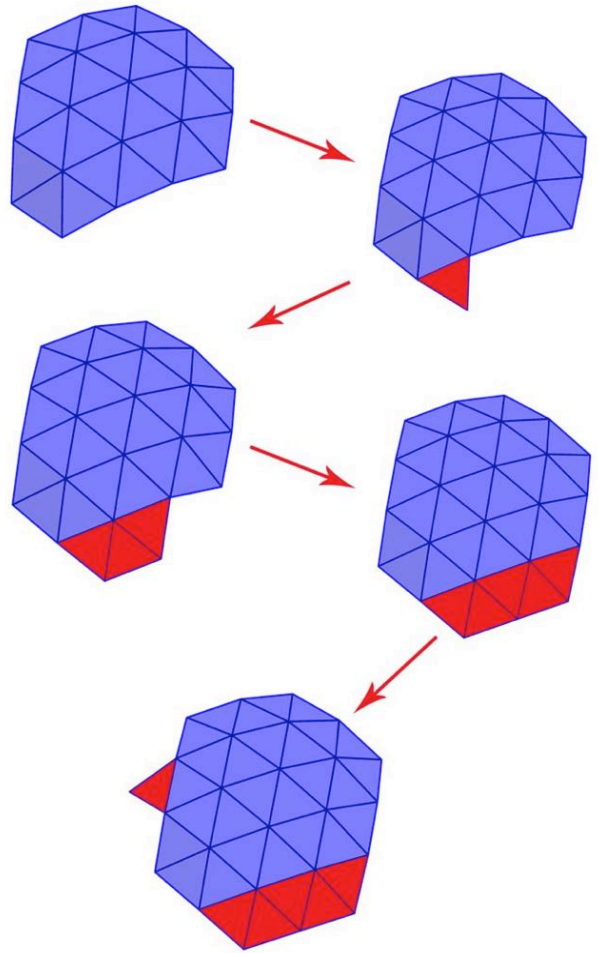

**Supplementary Figure 8**. (a) Formation of a pentamer or hexamer depends on the opening angle. Five assembled subunits can either become a pentamer or a new subunit can be added to them to become a hexamer. (b) The step by step growth of a shell by addition of triangular subunits.

Supplementary Table 1: Percentage of cones in total number of capsids (cones+cylinders).

| Number of CA subunits that initially transform from an immature lattice to the mature lattice without disassembly | Absence of membrane and genome | Presence of membrane but absence of genome | Presence of both membrane and genome |
|---|---|---|---|
| **disassembly/reassembly** | | | |
| 1 subunit | 37% | 52% | 83% |
| **displacive + reassembly** | | | |
| 24 subunits | | 53% | 88% |
| 100 subunits | | 53% | 89% |
| 400 subunits | | 0% | 100% |
| 600 subunits | | 0%* | 40%** |

*Rolling sheet.
**Some cones are defective.